%Paper: astro-ph/9504006
%From: SANDRA SAVAGLIO <SAVAGLIO@FIS.UNICAL.IT>
%Date: Tue, 4 Apr 1995 20:21:17 +0200 (WET-DST)

% This is PESO.DEM the demonstration file of
% the plain TeX PESO macro package from Springer-Verlag version 0.9

\def\lya{Ly$\alpha$~}
\def\lyb{Ly$\beta$~}
\def\etal{{\it et al.\/}~}
\def\kms{km~s$^{-1}$~}

\def\12{{1\ov 2}}

\def\ov{\over}

\input peso.cmm
\input epsf
\contribution{Abundances in the Lyman--alpha clouds}
\author{Sandra Savaglio@{1,2} and John Webb@2}
\address{@1Dipartimento di Fisica, Universit\`a della Calabria,
I--87036 Arcavata di Rende, Cosenza, Italy
@2School of Physics, University of New South Wales,
P.~O.~Box 1, Kensington, NSW 2033, Australia}

\abstract{We have re--examined the chemical composition of \lya clouds using
the composite--cloud technique, in which each \lya
line in a spectrum is shifted to its rest frame wavelength and all rest frame
spectra are co--added to form an `averaged' \lya cloud spectrum.
We illustrate how various estimates of the spectrum and redshift evolution
of the background ionizing UV flux lead to very different predictions for
the relative strengths of heavy element lines in the forest clouds.
We also show how the potential abundance limits depend on the various
observational quantities and how different procedures may be required for OVI
and CIV.
Preliminary results are presented from an analysis of two high redshift QSO
echelle spectra.}

\titlea{1}{Photoionization models of non--primordial \lya clouds}

The determination of the heavy element abundances in the Lyman forest clouds
is an important test of their origin.  The observational limitation is that,
for an
individual cloud, the expected heavy element column densities are low
and, generally below typical detection limits.  This difficulty
prompted Norris \etal (1983) to devise a technique in which
individual \lya lines are shifted to their
rest frame wavelengths and all rest frame spectra are co--added.
The most prominent heavy element transitions can then be searched
for in this high signal--to--noise averaged Lyman forest cloud spectrum,
yielding abundance measurements or upper limits.

However, the results derived in this way rely sensitively on knowing the
UV flux impinging on the clouds to a reasonable accuracy.
Since there is now substantially more information available concerning the
redshift evolution and spectral shape of the UV background ionizing
radiation field incident compared to earlier studies, we have
re--examined the topic.   In particular we have explored which species may
yield
the best constraints on heavy element abundances in the Lyman forest clouds.

Two quite different models for the redshift dependence of the intensity
of the UV flux at the Lyman limit $J_{912}$ have been adopted in this study,
representing a plausible range of possibilities.
In  case ($a$) we have used an estimate of  $J_{912}$ based on the
integrated UV flux from observed QSOs, modified by the expected
opacity of H and He associated with all QSO absorption systems (Madau 1992).
In case ($b$) we have combined several measurements of $J_{912}$
derived from the proximity effect over a wide range in redshifts
(Fig.~1).  In both cases we have adopted the spectral shape of the UV
background,
$J_{\nu}$, given by Madau (1992).

 \begfig 0cm
 \figure{1}{The two UV flux models used in the photoionization calculations.
  The solid
 line is our approximation to the observed proximity effect estimates.}
 \epsfxsize=7.6cm
 \epsfysize=6.5cm
 \hfil\epsffile{1.ps}\hfil
 \endfig

The photoionization code CLOUDY  (Ferland 1991) was used to compute ionic
column densities for an assumed heavy element abundance of [M/H] $= -2$
for both UV backgrounds illustrated in Fig.~1.  The
cloud HI column density used was $\log N$(HI) $=14$, the total hydrogen
volumetric
density was $\log n_H = -4$ and plane parallel slabs were assumed,
illuminated on one side by the UV background.  Scaled solar abundances are
adopted.  Results are shown in Fig.~2.  For
both UV models, OVI, CIV and NV are all prominent for $z>2$, but weak at lower
$z$
due to the lower UV background.
In Fig.~2($a$), at $z>2$, the UV flux and ion column densities remain almost
constant.
In Fig.~2($b$), the UV redshift dependence of the background radiation is
substantially
greater and for $z> 1.2$, the ion column densities change rapidly. At
redshift $z\sim3$, CIV and NV become more highly ionized
while OVI, which has the higher ionization potential, is less affected by the
rapidly evolving UV background flux.

 \begfig 0 cm
 \figure{2}{Heavy element abundances (X) relative to HI as function of redshift
 for the two different models for $J_{912}$: $a)$ from Madau (1992); $b)$
 from the proximity effect measurements.}
 \epsfxsize=10cm
 \epsfysize=5.5cm
 \hfil\epsffile{2.ps}\hfil
 \endfig

An interesting feature of Fig.~2 is the large variation between the
various column densities, particularly for the higher UV background model.
For UV background estimates based on the proximity effect,
OVI has the highest column density, being more than 500
times greater than that of CIV at $z \sim 3$ .  OVI however does of course
suffer from
confusion with the general forest distribution and thus it is not immediately
obvious that OVI can provide the best constraints.

\titlea{2}{ Potential Observational Constraints}

Here we describe some preliminary results from a limited exploration of
parameter--space using simulations of QSO spectra with various input
cha\-racteristics.
Synthetic QSO spectra were generated, using \lya cloud heavy element column
densities
based on the lower UV background radiation field model (i.e. the dashed line
of Fig.~1).   Here we have taken  [M/H] $= -2.5$, i.e. lower
abundances than those used in deriving Fig.~2.  The QSO emission redshift
is taken as 3.5, to consider \lya lines in the redshift range $2.8-3.5$.

 \begfig 0 cm
 \figure{3}{ Composite \lya forest OVI and CIV doublet spectra derived
 from simulated QSO spectra.
 The resolution  and S/N are fixed to 10  \kms and 10 respectively.
The right hand column indicates the number of spectra and the EW
threshold used in each case.}
 \epsfxsize=9cm
 \epsfysize=9cm
 \hfil\epsffile{figa5.ps}\hfil
 \endfig

In the initial set of simulations (top panel of Fig.~3), we used a spectral
resolution and signal--to--noise ratio typical of echelle spectrographs on
4m telescopes (FWHM = 10 \kms) and  S/N = 10 per pixel.
In generating the top set of composite spectra in Fig.~3
we computed 10 `observed' QSO spectra and varied the equivalent width (EW)
selection threshold from 0.5 \AA\ to 0.1 \AA.  On the left side of Fig.~3 the
region with the \lyb and the OVI doublet is shown and on the right hand side
the
CIV doublet.  The results suggest that if a low selection threshold is chosen
(0.1 \AA)
any OVI signal is diluted beyond detection (bottom curve).  On the other hand,
if the
threshold is rather high (0.5 \AA), the number of lines is small
and confusion noise due to line blending dominates (top curve).
The optimal value of the selection threshold for OVI constraints appears to be
close
to 0.3 \AA.  For CIV, where line blending with forest lines is not relevant,
the situation is
rather different and a selection threshold of EW $\geq 0.5$
\AA\ gives the strongest detection.

In the second set of simulations (bottom panel
of Fig.~3) the EW limit is fixed to 0.3 \AA\ and the number of spectra was
changed.
The results suggest that CIV could easily
be detected with just 5 spectra, whilst OVI needs at least 10, for the assumed
spectral characteristics.

Similar numerical experiments were carried out varying the S/N ratio and
spectral resolution.  The results indicate that,
as expected, the detectability of CIV is sensitive to S/N and thus 10m class
telescopes
are desirable if CIV is to be successfully used, whilst for the
confusion--limited
OVI the effect of increasing S/N is less marked.  Whilst CIV can be detected
even in relatively low resolution data, the detection of OVI
requires high resolution to overcome line confusion.

In the last example (Fig.~4) we have simulated spectra at different redshifts
for
constant abundances of $- 2.5$ and a signal--to--noise ratio of 10.
10 spectra were simulated at redshifts of 2.5, 3.5 and 4.5.  Fig.~4 suggests
that
CIV detection is relatively insensitive to redshift,
whilst OVI is only detected at $z=3.5$, presumably due to a more
favourable combination of the effects of
line confusion, the number of lines per unit redshift, and the
redshift dependence of the UV background.

 \begfig 0 cm
 \figure{4}{Composite \lya forest OVI and CIV doublet spectra derived
 from simulated QSO spectra with input abundances of [M/H] $=- 2.5$
The right hand column indicates the emission redshift, the spectral resolution,
the S/N and the number of spectra.}
 \epsfxsize=10.3cm
 \epsfysize=7cm
 \hfil\epsffile{5b.ps}\hfil
 \endfig

We emphasize that the results illustrated in Figs.~3 and 4 are based on the
{\it lower} UV background case. Fig.~2 shows that the results
would be substantially different if {\it proximity effect} UV estimates are
closer to
reality, in which case CIV would be extremely weak and provide no useful
constraints.  Interesting abundance constraints would then be derived only from
OVI.

Apart from one CIV study (Tytler \& Fan 1994), all previous work has been at a
spectral resolution of about 1 \AA~ or worse.
No high spectral resolution study of OVI has been done so far to our knowledge,
probably because the line confusion complicates matters and because even for
high redshift QSOs, the OVI forest still falls in a relatively blue spectral
region, where
many instruments/detectors are not efficient.
In Fig.~5 we illustrate preliminary results based on just 2 spectra, the $z =
4.1$ Q0000--26 and $z=4.5$ BRI1033--26.
Whilst the Ly$\beta$ line is clearly seen, no OVI is detected (the dashed line
in Fig.~5).  Upper limits on the OVI column density give upper limits on
the heavy element abundances of $- 1.5$ and $- 2$ for the two assumed
UV flux models (dashed and solid line of Fig.~1 respectively).
Further data are clearly required
to achieve lower limits or detections.

 \begfig 0 cm
 \figure{5}{Composite spectrum of the \lya forest for the
high redshift QSOs Q0000--26 and BRI1033--26 illustrating the Ly$\beta$ line
and OVI doublet.}
 \epsfxsize=9cm
 \epsfysize=7cm
 \hfil\epsffile{figa7.ps}\hfil
 \endfig

\begref{References}{}

\ref
Bechtold, J., 1994, ApJS, 91, 1.
\ref
Cristiani, S., D'Odorico, S., Fontana, A., Giallongo, E., Savaglio, S., 1995,
MNRAS, {\it in press}.
\ref
Ferland, G.J., 1991,  {\it OSU Astronomy Dept.~Internal Rept.}, 91--01.
\ref
Fernandez--Soto, A., Barcons, X., Carballo, R., Webb, J.K., 1995
MNRAS, {\it submitted}.

\ref
Giallongo, E., Cristiani, S., Fontana, A., Trevese, D., 1993, ApJ, 416, 137.
\ref
Kulkarni, V.P., Fall, S.M., 1993, ApJL, 413, L63.
\ref
Madau, P., 1992, ApJL, 389,L1.
\ref
Norris, J., Hartwick, F.D.A., Peterson, B.A., 1983, ApJ, 273, 450.
\ref
Tytler, D., Fan, X.--M., 1994, ApJL, 424, L87.
\ref
Williger, G.M., Baldwin, J.A., Carswell, R.F., Cooke, A.J., Hazard, C., Irwin,
M.J., McMahon, R.G., Storrie--Lombardi, L., 1994, ApJ, 428, 574.

\bye